\documentclass[useAMS,usenatbib]{mn2e}
\usepackage{graphicx}
\usepackage{times}
\usepackage[dvips]{color}

\def\mmm{$(m-M)_0$}
\def\ebv{$E(B-V)$~}

\def\vi{$V-I$}
\def\bv{$B-V$}

\def\msun{M$_{\odot}$}
\def\gsim{\;\lower.6ex\hbox{$\sim$}\kern-7.75pt\raise.65ex\hbox{$>$}\;}
\def\lsim{\;\lower.6ex\hbox{$\sim$}\kern-7.75pt\raise.65ex\hbox{$<$}\;}

\title[Berkeley\,17]{$BVI$ photometry of the very old open cluster 
Berkeley\,17\thanks{ 
Based on observations made with the Italian
Telescopio Nazionale Galileo (TNG) operated on the island of La Palma by
the Fundaci\'on Galileo Galilei of the INAF (Istituto Nazionale di
Astrofisica) at the Spanish Observatorio del Roque de los Muchachos of
the Instituto de Astrofisica de Canarias.}
}

\author[Bragaglia et al.]{Angela Bragaglia$^1$\thanks{E-mail: 
angela.bragaglia@oabo.inaf.it (AB), monica.tosi@oabo.inaf.it (MT),
andreuzzi@tng.iac.es (GA), gmarconi@eso.org (GM)},
 Monica Tosi$^1$, Gloria Andreuzzi$^{2,3}$ and Gianni Marconi$^{4}$\\
 \\
$^1$ INAF--Osservatorio Astronomico di Bologna, Via Ranzani 1, I-40127 Bologna
      (Italy) \\
$^2$  Fundaci\'on Galileo Galilei - INAF, Calle Alvarez de Abreu 70, 
      38700 Santa Cruz de La Palma, TF (Spain)\\
$^3$ INAF--Osservatorio Astronomico di Roma, Via dell'Osservatorio 5, 
      I-00040 Monte Porzio (Italy) \\
$^4$  ESO, Alonso de Cordova 3107, Vitacura, Santiago (Chile)
     }

\date{}

\begin{document}
\maketitle

\begin{abstract}
We have obtained $BVI$ CCD imaging of Berkeley~17,  an anticentre open
cluster that competes with NGC~6791 as the oldest known open cluster.
Using the synthetic 
colour magnitude diagrams (CMD) technique with three sets of evolutionary tracks
we have determined that its age is 8.5 - 9.0 Gyr, it distance modulus is
\mmm = 12.2, with a reddening of \ebv = 0.62 - 0.60.
Differential reddening, if present, is at the 5\% level.
All these values have been obtained using models with metallicity about half of 
solar (Z=0.008 or Z=0.01 depending on the stellar evolution tracks), which
allows us to reproduce the features of the cluster CMD better than other
metallicities.
Finally, from the analysis of a nearby comparison field we think to 
have intercepted a portion of the disrupting Canis Major dwarf galaxy.

\end{abstract}


\begin{figure}
\centering
\includegraphics[bb=10 10 260 670, clip,scale=0.75]{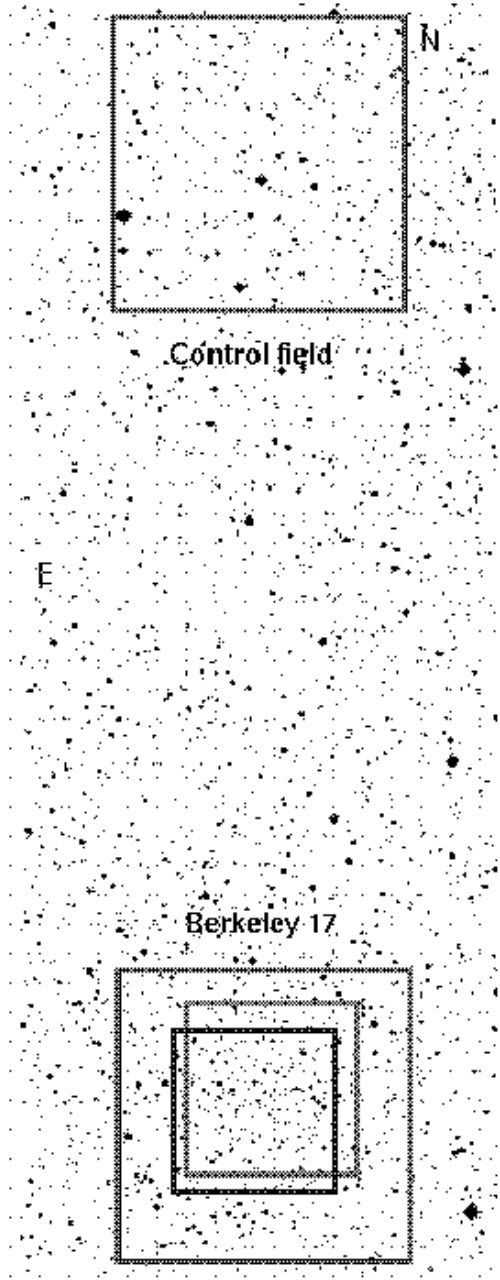} 
\caption{
Approximate positions of our pointings on Be~17 and the control field (larger
squares); the smaller squares indicate the positions of the central field in
K94 and P97. The map is 15 $\times$ 50 arcmin$^2$, has North to
the top and East to the left.	 
} 	 
\label{fig-map}
\end{figure}

\section{Introduction}

It is generally accepted that open clusters (OCs) are among the best
tracers of the properties, the formation and the evolution of the disc of
our Galaxy (e.g., \citealt{friel95}). This is particularly true for the
oldest ones, since they could trace back the first moments of the disc lifetime
and put crucial constraints on its formation mechanism(s) and early
evolution.

Berkeley~17 is among the oldest OCs of the Galaxy,  if not the oldest one. It
is located at $\alpha_{2000} = 05^h 20^m 32^s$, $\delta_{2000} = +30^o
33\arcmin 47\arcsec$, corresponding to the Galactic anticenter direction $l =
175.68$, $b = -3.68$. This cluster had not received much attention since its
discovery by \cite{sw62} until 1994, when the almost contemporary works by 
\cite{k94} (hereafter K94), \cite*{pjm94} and \cite{jp94} definitely found it
to be a very old object. The controversy on the actual age of Be~17  began
immediately, since  K94 stated that Be~17 is at least as old as NGC~6791, whose
age he assumed to be 9 Gyr, while \cite{pjm94} and \cite{jp94}  derived an age,
based on the Morphological Age Indicator (MAI, see Sect. 3) of about  12 Gyr,
i.e., similar to that of the youngest globular clusters  (albeit with a large
error bar).  A very large age has been reiterated by \cite{p97} (hereafter P97:
12$^{+1}_{-2}$ Gyr), but not  fully confirmed  by several other analyses (e.g.,
\citealt{carraro98a,carraro99}: $9 \pm 1$ Gyr; \citealt{salaris04}: $\simeq 10$
Gyr; \citealt{kc06}: $10 \pm 1$  Gyr).

The very large age found by P97 would imply that the age distributions of open
and globular clusters overlap, and that the disc began to form stars during the
last phases of halo formation. On the contrary, if Be~17 is conclusively
demonstrated to be younger than the youngest globular clusters (GCs),  
the existence
of a hiatus between the formation of the halo/thick disc and the present thin
disc  would be reinforced. The latter is nowadays the most popular scenario
for the formation of our Galaxy and is supported by many 
observational evidences \citep{fbh02}.

Less dispute appears to exist about the metallicity of Be~17, that has been
estimated to be slightly lower than solar by \cite{friel02}, with 
[Fe/H]=$-0.33~(\sigma=0.12)$ dex, based on low resolution spectroscopy, and by
\cite{friel05}, with [Fe/H]=$-0.10~(\sigma=0.09)$ dex, based on high
resolution spectroscopy of three giant stars. A slightly sub solar
or at most nearly solar abundance is also found by photometric analyses,
both in the optical and IR bands \citep[see. e.g., P97;][]{carraro98a,carraro99}.

Even if its age is reduced to a value of 9-10 Gyr,
Be~17 is a fundamental cluster because it defines the old tail of the OC 
population together with NGC~6791, from which it differs significantly.
We are building a sample of homogeneously studied OCs
well distributed in age, position and metallicity, that we intend to use to
study the disc properties and evolution;
see \cite{bt06} for a detailed description of our program (that we named
Bologna Open Cluster Chemical Evolution - or BOCCE - program) and its
results. Such a sample cannot leave Be17 aside.
We have obtained
deep $B,V,I$ photometry that we present here and high resolution spectra of
3 red clump stars that will be the subject of a forthcoming paper.
Medium resolution spectra of stars from the tip of the red giant branch to the
main sequence are planned to determine membership of more than 100
stars in this crucial cluster.

The paper is organized as follows: we present our data 
in Section 2, we discuss the resulting CMDs in Section 3 and derive the
fundamental cluster parameters using the synthetic CMDs method in Section 4;
finally a discussion and a summary are given in Section 5.

\section{Observations and data reduction}

Observations of Be\,17 and of the associated control field used to check
the field stars contamination were obtained at the Telescopio Nazionale 
Galileo on UT 2000 October 1, 2, and 3. We
used DOLORES (Device Optimized for the LOw RESolution), a focal reducer capable
of imaging and low resolution spectroscopy,  mounting a 2k Loral thinned and
back-illuminated CCD, with scale of 0.275 arcsec/pix, and a field of view
9.4 $\times$ 9.4 arcmin$^2$. The two fields, separated by about 30
arcmin, are shown in Fig.~\ref{fig-map}. 
The observations include several exposures in each of the three
$BVI$ Johnson-Cousins' filters. Table 1 lists the dates of the 
observations together with the filters used and the corresponding ranges of 
exposure time (in seconds).
All the three nights were photometric, with a mean seeing value around 
1.4$\arcsec$.

\begin{table*}
\begin{center}
\caption{Log of observations for the cluster ($\alpha_{2000}$ = 
05$^{h}$ 20$^{m}$ 32$^{s}$; $\delta_{2000}$ = 30$^{0}$ 33' 47''; $l=175.68$,
$b=-3.68$)
and the external field ($\alpha_{2000}$ =05$^{h}$ 20$^{m}$ 32$^{s}$; 
$\delta_{2000}$ = 31$^{0}$ 03' 47''; $l=175.26$, $b=-3.40$). 
N$_V$, N$_B$ and N$_I$ are the number of exposures ($>$ 1) in each filter; 
t is the exposure time in seconds.}
\vspace{5mm}
\begin{tabular}{l|lll}
\hline\hline
field &  N$_{V}$ $\times$ t & N$_{B}$ $\times$ t & N$_{I}$ $\times$ t\\
\hline 
Cluster &   3$\times$600, 200,  60,  30, 10,  5 & 1800, 2$\times$900, 2$\times$600, 60  & 471, 400, 4$\times$300, 120, 2$\times$20, 10, 5\\
External  & 2$\times$300, 10, 2 & 900, 300, 60,  10, 2 & 2$\times$300, 10, 1 \\
\hline
\end{tabular}
\end{center}
\label{tab-oss}
\end{table*}

Corrections for bias and flat-field were performed using the standard 
IRAF\footnote{IRAF is distributed by the National Optical Astronomical
Observatory, which are operated by the Association of Universities for
Research in Astronomy, under contract with the National Science Foundation }
procedures. The subsequent data reduction and analysis was done following  the
same procedure for both data-sets (cluster and comparison field), and using
the {\sc DAOPHOT ii}  package in IRAF \citep{stetson87,davis94}.  Objects have 
been independently searched in all the frames, using  a threshold of 4
$\sigma$ above the background, and have been measured with a quadratically
varying point spread function (PSF).    For each filter,  the magnitudes have
been aligned to that of a reference frame, the deepest one, obtained
in the best seeing conditions.  An average instrumental magnitude (weighted
with the photometric errors) was derived in each filter. 
For the brightest stars, which are saturated in all frames but the shortest
ones, only one measure was available. 
Finally, we applied aperture corrections (of the order of 0.2 - 0.3 mag) to the   
weighted instrumental PSF magnitudes in each band, to calibrate them on the same    
system of the standard stars. For the latter, magnitudes were
derived using aperture photometry. 

The final catalogs have been created including all the objects identified in
at least two filters, after applying a moderate selection in the shape-defining
parameter $sharpness$  ($-2 \leq sharpness \leq 2$) and on the goodness-of-fit
estimator $\chi^2$   ($\chi^2 \leq 10$). To the two final catalogs, one for the 
cluster and one for the comparison field, we applied the transformation to
astrometrize the $\alpha$ and $\delta$ coordinates, using software written by P. Montegriffo at the
Bologna Observatory.

\begin{figure}
\begin{center}
\includegraphics[bb=15 160 565 690, clip, scale=0.44]{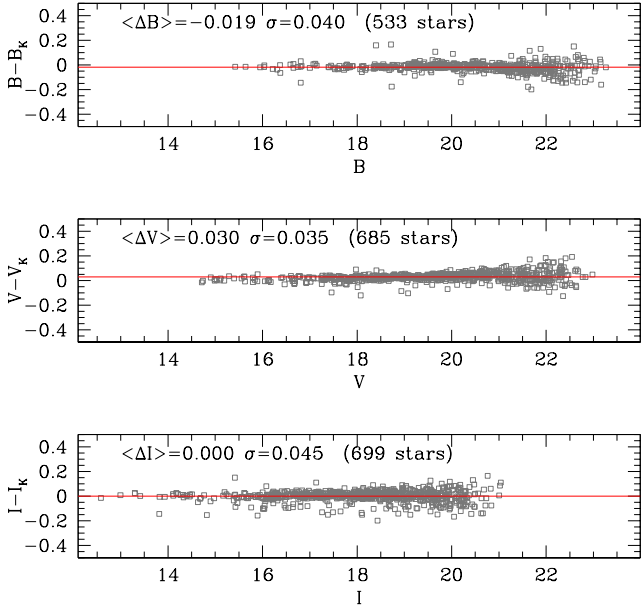} 
\caption{Comparison between our photometry and the one by K94}	
\label{fig-confk}
\end{center}
\end{figure}

\begin{figure}
\begin{center}
\includegraphics[bb=15 160 565 690, clip, scale=0.44]{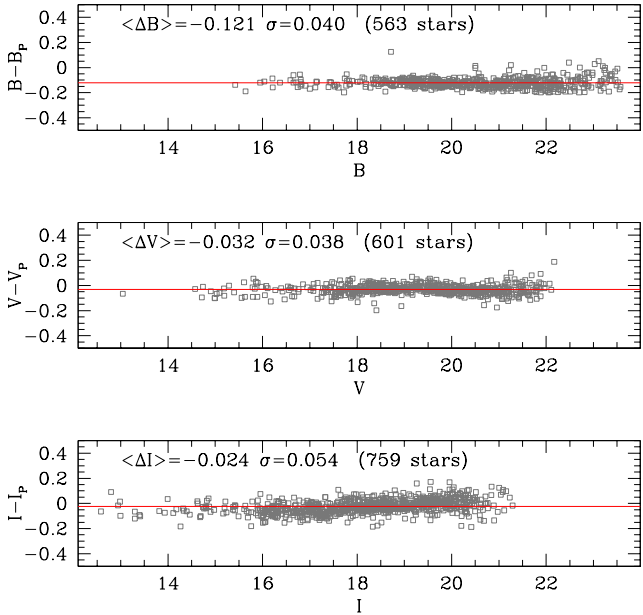} 
\caption{Comparison between our photometry and the one by P97}	
\label{fig-confp}
\end{center}
\end{figure}

\subsection{Photometric calibration}

Transformation from the instrumental to the standard system have been obtained
using the standards areas PG0231+051 and MarkA (Landolt 1992), observed several
times during the three nights. 
We derived the calibration equations using only the 
standard fields observed on the same nights of the reference frames (October 2
and 3). 
The 10 stars retained in our calibrations have colours
$-0.329 <B-V< 1.448$ and $-0.534 <V-I< 1.951$, thus covering
the bulk of the cluster stars without need for extrapolation.

For the extinction coefficients we used the average between the values  
of these two nights\footnote {see
www.ast.cam.ac.uk/$\sim$dwe/SRF/camc\_extinction.html  and 
\cite{king85}} 
($\kappa_B=0.23, \kappa_V=0.13, \kappa_I=0.03$). The resulting calibration
equations have the following form:

\[  B = b +0.0460 \times (b-v) +1.4091  ~~(rms=0.013) \]
\[  V = v -0.1864 \times (b-v) +1.2310  ~~(rms=0.011) \]
\[  V = v -0.0936 \times (v-i) +1.1895  ~~(rms=0.016) \]
\[  I = i +0.0419 \times (v-i) +0.7143  ~~(rms=0.013) \]

\noindent
where $b, v, i$, are the aperture corrected instrumental magnitudes, after
correction also for extinction and exposure time, and $B, V, I$ are the
output magnitudes, calibrated to the Johnson-Cousins standard system. 

Finally, we determined our completeness level in each band using extensive
artificial stars experiments: we iteratively added, one at a time, about 50000 simulated stars
to the deepest frames and repeated the reduction procedure, determining the
ratio of recovered over added stars (see \citealt{tosi04} for a more detailed
description). The results are given in Table 2.

We checked the calibration comparing our photometry with that presented
in previous  literature (i.e., K94 and P97, which supersedes the one by
\citealt{pjm94}, taken in non optimal weather conditions). To estimate
the differences, we cross identified our objects 
with those of K94 and P97 (obtained through the BDA\footnote{
http://www.univie.ac.at/webda/new.html}, \citealt{mermio95}). 
We found (see Figs \ref{fig-confk} and
\ref{fig-confp}):
$\langle \Delta B_K\rangle = -0.019$, $\langle \Delta V_K\rangle  = +0.030$, 
$\langle \Delta I_K \rangle = 0.000$ for K94,
where $\Delta$ is intended as our magnitude minus the one by K94, and
$\langle \Delta B_P \rangle = -0.121$, $\langle \Delta V_P \rangle = -0.032$, 
$\langle \Delta I_P \rangle =-0.024$
for P97. Even if in neither case there are strong trends with magnitude,
the comparison looks better with K94. The differences
are small, except for our B photometry compared to P97's. 
The nights in which we observed the reference frames were deemed photometric,
as was the case for the P97 observations (while K94 used the photometric
part of a not completely photometric night), so we cannot easily explain
this large difference.

These differences in magnitudes translate into differences in
colours:
$\langle \Delta(B-V)_K \rangle = -0.049$, $\langle \Delta(V-I)_K\rangle = +0.030$ and
$\langle \Delta(B-V)_P\rangle = -0.089$, $\langle \Delta(V-I)_P\rangle = -0.008$: we are bluer than both
literature photometries in the $V,B-V$ plane; redder than K94 and almost
identical to P97 in the
$V,V-I$ plane. Since the simultaneous fit of the $B-V$ and $V-I$ colours is one of
the indicators to select the most appropriate metallicity from the comparison
with theoretical stellar models (see Section 4), such an uneven difference in
the colours may affect the photometric metallicity choice.

We also compared the two photometries by K94 and P97 with each other, 
finding differences in agreement with what is given above; if 
we limit the comparison to stars brighter than
$V = 20$ as in P97 (their table 1), we find:
$\langle  B_P - B_K \rangle= 0.116 $,
$\langle  V_P - V_K \rangle= 0.060 $, and
$\langle  I_P - I_K \rangle= 0.058 $, based on about 280 stars in common.
This means 
$\langle \Delta(B-V)_{P-K} \rangle = 0.056$, 
$\langle \Delta(V-I)_{P-K} \rangle = 0.002$.
As already stated by P97, there are no direct means to decide which photometry is
more precisely on the standard system (there are no stars with photoelectric measures
in our fields), so we will proceed with our own data, without attempting
any correction.

After all these comparisons were made, a new photometry has been presented
(but is not yet available) by \cite{kc06}.  It appears of slightly lower quality
than ours (e.g., the cluster sequences are less well defined) but, thanks to
differential comparison with NGC 188,   it is claimed
to be on the standard Landolt system to within $\pm$ 0.03 mag. Interestingly,
they too find their photometry bluer than the one by P97 by
about 0.1 mag in ($B-V$), providing further support to our calibration.

\begin{figure*}
\includegraphics[bb=10 175 575 650, clip,scale=0.6]{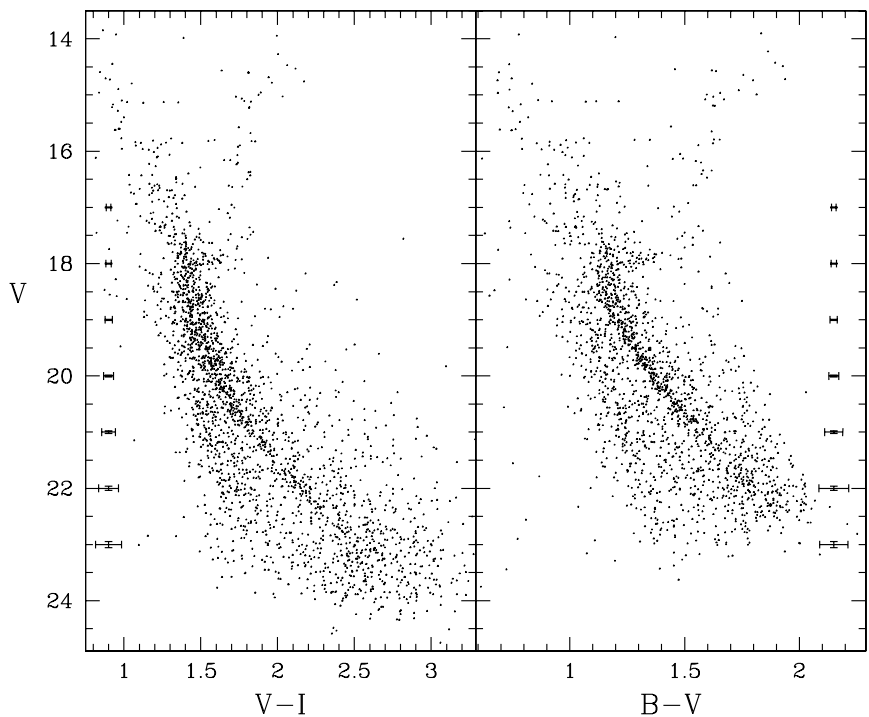} 
\caption{Left panel: $V, V-I$ CMD for Be\,17; Right panel: $V, B-V$ CMD.
	The mean errors per interval of magnitude V are also plotted.
	}	
\label{fig-cmd}
\end{figure*}

\begin{figure*}
\includegraphics[bb=10 175 575 650, clip,scale=0.6]{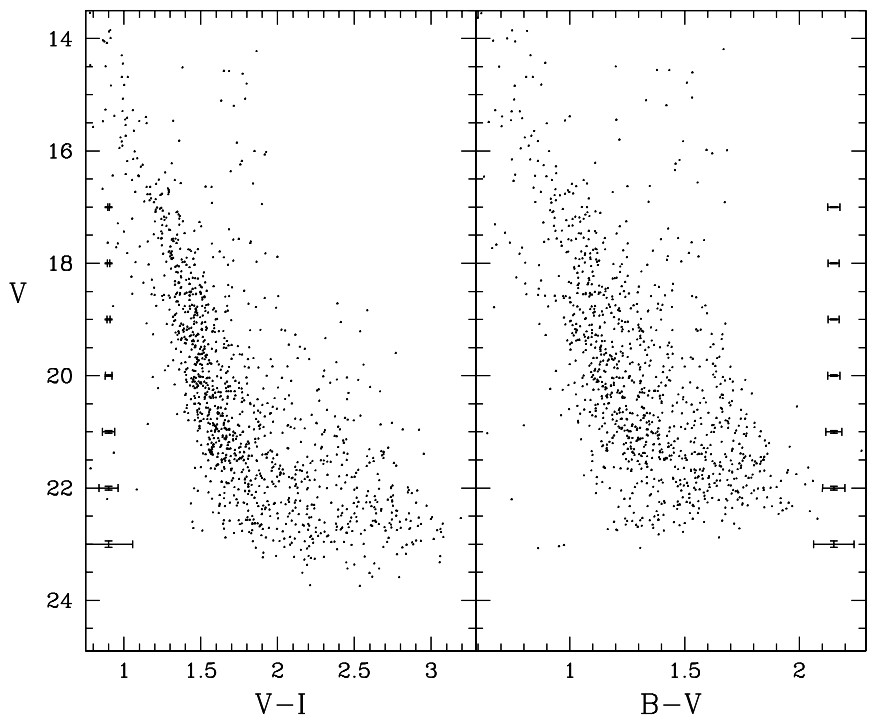} 
\caption{Left panel: $V, V-I$ CMD for the external field; Right panel: $V, B-V$
        CMD.
        The mean errors per interval of magnitude V are also plotted.
	} 
\label{fig-cmde}
\end{figure*}

\begin{figure*}
\begin{center}
\includegraphics[scale=0.7]{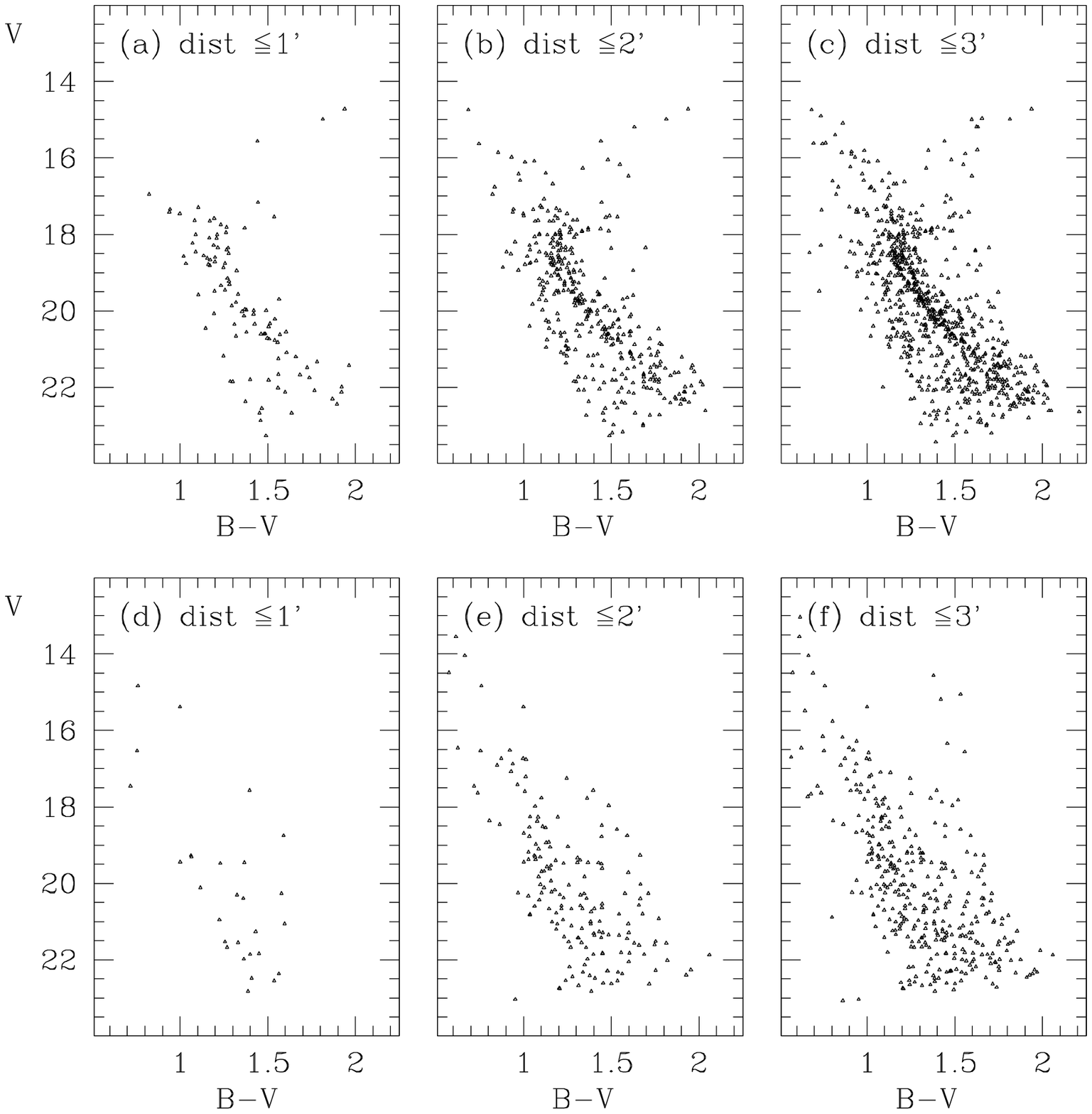} 
\caption{
Radial CMDs for Be~17 (upper panels) and corresponding areas of the
comparison field (lower panels); the center was assumed at pixel x=1225, y=1175
in both cases, and we plot stars within distances of 1, 2, 3 arcmin from it.
The CMDs contains 103, 385, 790 objects for panels (a), (b), (c) respectively,
and 28, 177, 372 for panels (d), (e), (f) respectively.}
\label{fig-rad}
\end{center}
\end{figure*}

\section{The colour - magnitude diagram}

The final, calibrated sample of the cluster stars (that will be made available
through the BDA) consists of 2473 objects identified in  at least two filters,
of which 1940 identified simultaneously in all the three filters $BVI$.

The corresponding CMDs are shown in Fig. \ref{fig-cmd}, where the mean errors 
per  magnitude bin are also plotted.  Both CMDs show: 
(i) a very clear main sequence (MS) extending down to V $\sim$ 24 for the $V,
V-I$ CMD and to  V $\sim$ 23  for the $V,B-V$ CMD; 
(ii) a main sequence turn-off (MSTO) near  V = 18, with a  sparsely
populated red giant branch (RGB), extending more than three magnitudes above the
TO. This morfology is typical of very old stellar systems for which the RGB
raises very steeply and is very extended vertically, while the colour separation
between the MSTO and the base of the RGB is smaller than in intermediate age
clusters.
The red clump (RC), including $\sim$ 8 stars, is visible at $V \sim 15$ and
$V-I \sim 1.7$, $B-V \sim 1.6$. This clump coincides with that identified by
K94 and P97 and, in our opinion, is also apparent in \cite{kc06} CMD, in spite
of these authors' conclusion.

We have measured the difference in magnitude between the RC and the MSTO:
$\delta V \simeq 2.9 \pm 0.1$. This, using the so called MAI
\citep{jp94}, implies an approximate age of 14.5 Gyr, but even the
conservative error of 0.1 mag in the $\delta V$ value means that the cluster
could be more than 16 Gyr old, or as "young" as about
12 Gyr. Since it is now generally accepted that this calibration leads to ages
systematically larger than those derived e.g, from isochrone fitting,  
we take this only as an indication of the very old age of the cluster.
We intend to re-calibrate the age-$\delta V$ relation  
once the entire sample of OCs in our BOCCE program is examined and
homogeneously dated through stellar evolutionary models. 
 
In both CMDs the sequence of binary stars is visible, although its
identification is complicated by the presence of a substantial field star
contamination. For this reason, we did not attempt to quantify the fraction of
binary systems as we did for less contaminated clusters (e.g., NGC~6253: 
\citealt{bragaglia97}; Be~21: \citealt{tosi98}). 

Finally, we note that information on membership based on radial velocity is
available only for  a few bright stars. \cite{scott95} observed about 30 stars
(all brighter than $V \sim 16$)   determining velocities with a precision of
about $\pm$ 10 km~s$^{-1}$ and their data were used by \cite{friel02} for the
determination of the cluster metallicities. They found that only about half of
them could be considered good candidate cluster members. They measured a mean
velocity  of $-84$ ($\sigma$ = 11) km~s$^{-1}$ and a mean metallicity  [Fe/H]=$
-0.33$ ($\sigma$ = 0.12) dex. Four of these stars were later studied by
\cite{friel05}, who find that only three are actual cluster members and derive
an average velocity of $-73.7$ ($\sigma$ = 0.8) km~s$^{-1}$ and a mean
[Fe/H]=$-0.10$ ($\sigma$ = 0.09) dex.

\begin{figure*}
\begin{center}
\includegraphics[bb=30 180 580 480, clip, scale=.9]{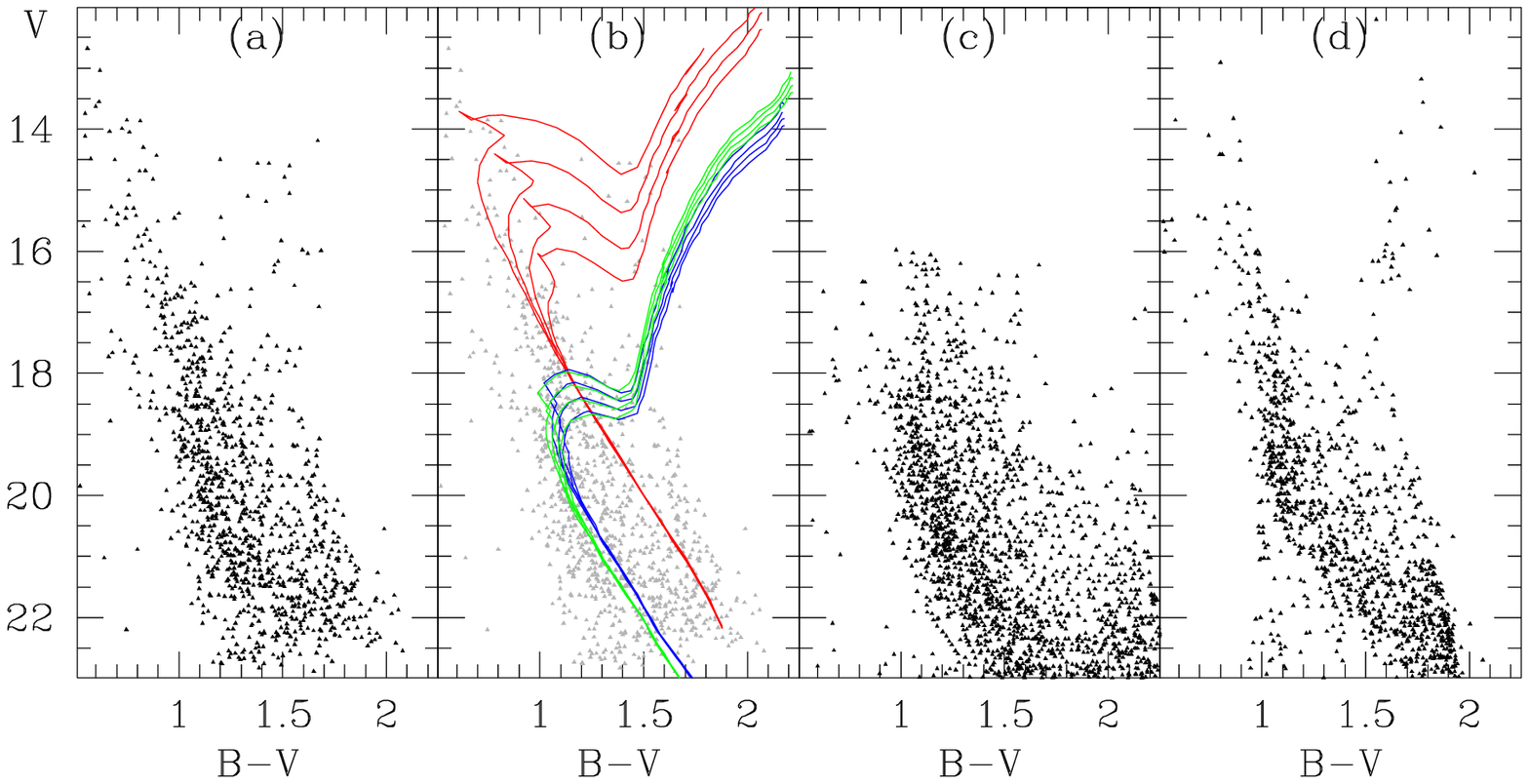} 
\caption{What kind of pupulation is present in the comparison field (panel a) ?
In panel (b) we show that it can be the combination of a young, nearby
component, with \ebv$\sim$0.5,  \mmm$\sim$11.75), corresponding to
the Perseus arm, and an older, farther, metal-poorer one. The reddening and 
distance [\ebv$\sim$0.60-0.65, \mmm$\sim$13.30-13.35] of the latter suggest 
that we could have
intercepted the orbit of the disrupting CMa. The younger population is fitted
with the four brighter dark isochrones (red in the colour electronic version),
with age 0.8, 1.25, 2 and 3 Gyr. 
The older population is compared with the four lighter (green in the colour
version) Z=0.004 isochrones, and with the four dark (blue) Z=0.008 ones,
corresponding in both cases to ages of 4, 5, 6.3 and 9.9 Gyr.
In panel (c) we plot the CMa CMD
\citep{bellazzini04} assuming the reddening and distance given above
[after correction for the values of \ebv=0.10, \mmm=14.5 in
\citealt{bellazzini04}] and scaling for the different areas. 
Panel (d) shows the result of the Galactic model by \citet{robin03} in the
direction of our field, after correction for the completeness function.
}
\label{fig-ring}
\end{center}
\end{figure*}

\subsection{The control field}

To estimate the field stars contamination we used the data of the control field,
 which is  far enough from the cluster to be safely considered free of
cluster members. Fig. \ref{fig-cmde} shows the
resulting $V, V-I$ and $V, B-V$ CMDs for the objects identified in this field. 
The corresponding 
catalogue contains 1310 objects identified in at least two
filters, of which 1103 are identified simultaneously in all the three filters.
Comparison of Figs. \ref{fig-cmd} and \ref{fig-cmde} helps to understand
which are the true cluster features. We also plot (see Fig. \ref{fig-rad})
radial CMDs, both for Be~17 (where the centre was defined using histograms 
of the star coordinates along the x and y axes) and the comparison field.
The cluster is quite loose, but its features clearly stand out with respect
to the field population in the central part.

As in many other cases (see e.g., \citealt{dorazi06} for Be~32), 
the CMD of the external field appears to be composed by (at least) two
populations. As already noted by K94, there is a broad main sequence that 
crosses diagonally the diagram and can be explained by the young population of
the Perseus spiral arm. As a demonstration, in Fig. \ref{fig-ring}(b) we have
plotted the Z=0.02 isochrones by \cite{bertelli94} for ages of 0.80, 1.25, 2,
and 3 Gyr, assuming the values usually attributed to the arm \ebv=0.5 and \mmm=11.75 (i.e., a distance from the Sun of
about 2.2 kpc). 

There is a second sequence,  fainter and bluer than the cluster MS (noticeable
also in the CMD of the  field centered on Be~17).  In Fig. \ref{fig-ring}(b) we
have overplotted on this second sequence the Z=0.008 and Z=0.004 isochrones
with older ages  (4, 5, 6.3, 9.9 Gyr) and assuming larger  distance and
reddening. We find a reasonably good fit with \ebv $\sim$ 0.65, \mmm $\sim$
13.35, i.e., a distance of about 4.7 kpc, if Z=0.004 and \ebv $\sim$ 0.60, \mmm
$\sim$ 13.30, i.e., a distance of about 4.8 kpc, if Z=0.008.  With these
values, we cannot have intercepted the Monoceros (or Anticentre) Ring, a
feature of the Galactic disc  visible towards the external parts of our Galaxy
(e.g., \citealt{newberg02}, \citealt{ibata03}) but with a much larger distance
from the Sun, of about 10 kpc. This Ring has been associated
(\citealt{martin04}, but see \citealt{momany04} for a different view) to an
over density in star counts, interpreted as  the remnant of the dwarf galaxy
Canis Major (CMa).
Interestingly, \cite{bellazzini06}  present the N-body model
of the disruption of CMa, which shows a clear signature towards the anticentre,
and at a distance of about 5 kpc from the Sun (see their fig. 11). We are possibly seeing this
feature, so we have compared the CMD of CMa to the one of our control field. We
took the data presented by \cite{bellazzini04} for CMa, taking into  account
the different areas, as well as the distance modulus and reddening given in
that paper, i.e. \ebv = 0.10 and \mmm = 14.5. The result is shown in
Fig.\ref{fig-ring}(c); the two CMDs are comparable, and we also see the blue
plume (for $V$ magnitudes between about 17.5 and 19, and colours bluer than
about $B-V$ = 1). These comparisons suggest that we are actually seeing a small
portion of the disrupting dwarf. 

Finally, Fig.\ref{fig-ring}(d) shows the output of the Besan\c con Galactic
model \citep{robin03} in the direction of our control field, after correction
for our completeness function; the bright parts of the CMDs are similar, but
they are a poor match fainter than about $V=20$. This shows that we cannot
explain our observed features in terms of simple Galactic field, but we need to
take the {\it anomalous} overdensities into account to reproduce all the
sequences of the CMD.

\begin{table}
\begin{center}
\caption{Completeness level for the central and external fields; mag is
the calibrated magnitude ($B, V$ or $I$, calibrated with the equations given
in the text, and assuming $b-v$ and $v-i$ = 1).  
}
\begin{tabular}{clllclll}
\hline\hline
   mag  & c$_B$  &c$_V$    &c$_I$   & &c$_B$   &c$_V$	 &c$_I$  \\
\hline
  16.00 &   1.0  &   1.0   &  1.0   & &   1.00 &   1.00  &  1.00 \\
  16.50 &   1.0  &   1.0   &  0.97  & &   1.00 &   1.00  &  1.00 \\
  17.00 &   1.0  &   1.0   &  0.99  & &   1.00 &   1.00  &  1.00 \\
  17.50 &   1.0  &   1.0   &  0.96  & &   1.00 &   1.00  &  1.00 \\
  18.00 &   1.0  &   0.97  &  0.98  & &   1.00 &   1.00  &  0.98 \\
  18.50 &   1.0  &   0.98  &  0.96  & &   1.00 &   1.00  &  0.99 \\
  19.00 &   0.99 &   0.99  &  0.94  & &   0.98 &   0.98  &  0.96 \\
  19.50 &   0.99 &   0.97  &  0.88  & &   0.97 &   1.00  &  0.96 \\
  20.00 &   0.98 &   0.97  &  0.81  & &   0.98 &   0.97  &  0.92 \\
  20.50 &   0.98 &   0.97  &  0.63  & &   0.96 &   0.97  &  0.80 \\
  21.00 &   0.97 &   0.95  &  0.31  & &   0.96 &   0.95  &  0.48 \\
  21.50 &   0.94 &   0.93  &  0.13  & &   0.92 &   0.97  &  0.12 \\
  22.00 &   0.94 &   0.91  &  0.03  & &   0.89 &   0.93  &  0.02 \\
  22.50 &   0.91 &   0.73  &  0.0   & &   0.87 &   0.87  &  0.0  \\
  23.00 &   0.81 &   0.45  &  0.0   & &   0.66 &   0.55  &  0.0  \\
  23.50 &   0.57 &   0.18  &  0.0   & &   0.38 &   0.29  &  0.0  \\
  24.00 &   0.24 &   0.04  &  0.0   & &   0.13 &   0.04  &  0.0  \\
  24.50 &   0.05 &   0.0   &  0.0   & &   0.02 &   0.0   &  0.0  \\
  25.00 &   0.0  &   0.0   &  0.0   & &   0.00 &   0.0   &  0.0  \\
\hline
\end{tabular}
\label{compl}
\end{center}
\end{table}

\section{Cluster parameters}

Age, distance and reddening of Be\,17 have been derived with the same
procedure  applied to all the clusters of our BOCCE project (see \citealt{bt06}
and references    therein), namely the synthetic CMD method originally
described by \cite{tosi91}. The best values of the parameters are found by
selecting the cases providing synthetic CMDs with morphology, colours, number
of stars in the various evolutionary phases and luminosity functions (LFs)
in better agreement with the observational ones.

As usual, to test the effects of the adopted stellar evolution models on the
derived   parameters, we have run the simulations with three different types of
stellar    tracks, with various prescriptions for the treatment of  convection
and of overshooting from convective regions.  Actually, Be\,17 is so old that
its MSTO stars have masses    around 1 M$_{\odot}$ and are therefore not
expected to experience any  overshooting from convective cores. Indeed, we find
no appreciable difference in the results obtained with models with or without
overshooting.

To estimate the metallicity which better reproduces the photometric properties 
of the cluster, we have created the 
synthetic CMDs adopting, for each type of stellar models, metallicities 
ranging from solar down to 20\% of solar. We still assume (see          
\citealt{bt06}) as
solar metallicity models those with Z=0.02, because they are the ones 
calibrated by their authors on the Sun, independently of the circumstance  that
nowadays the actual solar metallicity is supposed to be lower (see 
\citealt{asp05}). At any rate, we consider  the metallicities     obtained
with  our photometric  studies only indicative and use, whenever available,
high resolution spectroscopy for a safe determination of the chemical
abundances.

The adopted sets of stellar tracks are listed in Table 3, where the 
corresponding references are also given, as well as the  model metallicity and
the information on their corresponding overshooting assumptions. 
The transformations from the theoretical luminosity and effective temperature 
to the Johnson-Cousins magnitudes and colours have been performed using Bessel, 
Castelli \& Pletz (1998) conversion tables and assuming $E(V-I)$ = 1.25 \ebv
(Dean et al. 1978) for all sets of models. Hence, the different results
obtained with different stellar models must be ascribed fully to the models
themselves and not to the photometric conversions.

\begin{table}
\begin{center}
\caption{Stellar evolution models adopted for the synthetic CMDs. The FST
models actually adopted here are an updated version of the published ones
(Ventura, private communication)}
\vspace{5mm}
\begin{tabular}{cccl}
\hline\hline
   Set  &metallicity & overshooting & Reference \\
\hline
BBC & 0.02 & yes &Bressan et al. 1993 \\
BBC & 0.008& yes &Fagotto et al. 1994 \\
BBC & 0.004& yes &Fagotto et al. 1994 \\
FRA & 0.02 & no &Dominguez et al. 1999 \\
FRA & 0.01 & no &Dominguez et al. 1999 \\
FRA & 0.006 & no &Dominguez et al. 1999 \\
FST & 0.02 & $\eta$=0.00 &Ventura et al. 1998\\
FST & 0.02 & $\eta$=0.02 &Ventura et al. 1998\\
FST & 0.02 & $\eta$=0.03 &Ventura et al. 1998\\
FST & 0.01 & $\eta$=0.00 &Ventura et al. 1998\\
FST & 0.01 & $\eta$=0.02 &Ventura et al. 1998\\
FST & 0.01 & $\eta$=0.03 &Ventura et al. 1998\\
FST & 0.006 & $\eta$=0.00 &Ventura et al. 1998\\
FST & 0.006 & $\eta$=0.02 &Ventura et al. 1998\\
FST & 0.006 & $\eta$=0.03 &Ventura et al. 1998\\
\hline
\end{tabular}
\end{center}
\label{models}
\end{table}

The synthetic stars are attributed the photometric error derived from  the
artificial stars tests performed on the actual images. They are retained in (or
excluded from) the synthetic CMD according to the photometry  completeness 
factors listed in Table 2. All the synthetic CMDs have been computed either
assuming that all the cluster stars are single objects or that a fraction of
them are members of binary  systems with random mass ratio (see \citealt{bt06}
for a description of how binaries are included in the synthetic CMDs). We find,
as in many other clusters, that a  binary fraction around 30\% well reproduces
the observed distribution along the main sequence. All the synthetic CMDs shown
in the figures assume this  fraction of binaries. 

In spite of the field contamination, the evolutionary sequences are quite well
defined in the CMD of Be\,17. Fig. \ref{fig-rad} shows that the CMDs of the
inner cluster regions are not much cleaner than the global one, and have the
disadvantage of containing less stars; we have thus run the simulations for the
whole field covered by our images. Since the cluster field contains 1940 stars
measured in $B, V, I$ and the external field 1103, we assume that the cluster
members are 1940 -- 1103 = 837. Hence, the synthetic CMDs have been created
with 837 objects.

We find that in all cases a solar metallicity must be excluded, because it
does not allow us to simultaneously reproduce both the observed \bv \ and \vi \
colours. With Z=0.02, when the synthetic \bv \ is correct, \vi \ always turns
out  bluer than observed. Vice versa, models with Z$<$0.006 provide \vi \
colours redder than observed, when \bv \ is correct. On the other hand, 
stellar tracks with metallicity about half of solar can reproduce very well 
the observed colours of all the evolutionary phases, once the appropriate 
reddening is adopted. For the BBC models, the only available metallicity 
leading to self-consistent results is Z=0.008. For both the FRA and the FST
models, Z=0.006 allows to reproduce rather well the observed colours, but
Z=0.01 leads to a much better agreement. We thus favour Z=0.01 as the {\it
photometric} metallicity of Be\,17.

\begin{figure}
\centering
\includegraphics[bb=160 180 320 715, clip]{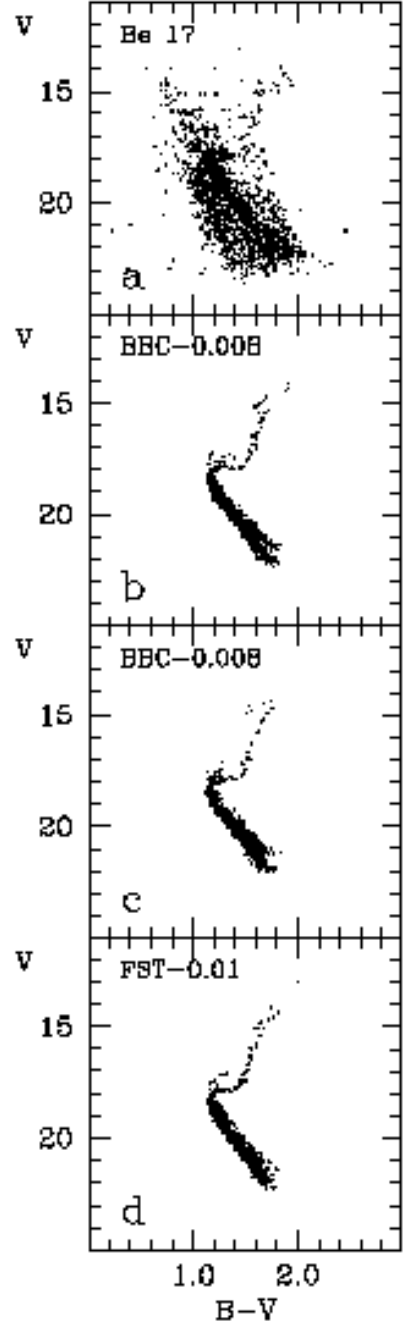}  
\caption{Comparison between empirical and synthetic CMDs. Panel a shows the \bv
\  CMD of the 1940 stars with measured $B, V$ and $I$. Panels b, c and d show
representative cases of synthetic diagrams of the 837 assumed cluster stars.
Panel b illustrates the best case for the BBC models: it assumes Z=0.008, 
age=8.5 Gyr, \ebv=0.62, and \mmm=12.2. Panel c shows the BBC predictions for an
age of 12 Gyr: the best metallicity is still 0.008, \ebv becomes 0.56 and 
\mmm= 12.1. Panel d illustrates the best case for the FST models: Z=0.01, 
age=9 Gyr, \ebv=0.60 and \mmm=12.2.} 
\label{simbv}
\end{figure}

\begin{figure}
\centering
\includegraphics[bb=160 180 320 715, clip]{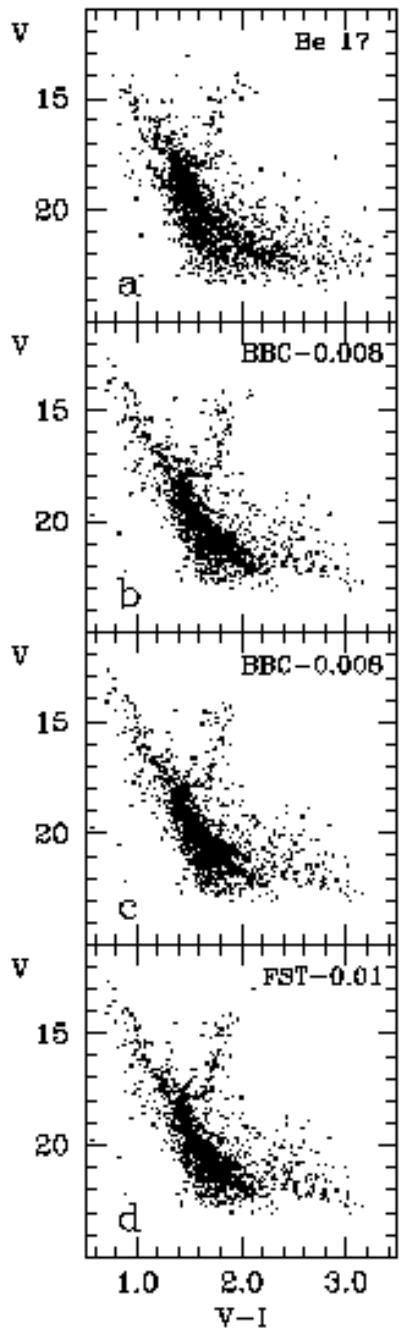}  
\caption{Comparison between empirical and synthetic CMDs. Panel a shows the \vi
\ CMD of the 1940 stars with measured $B, V$ and $I$. Panels b, c and d show
the same representative cases of synthetic diagrams of Fig. \ref{simbv}, but
with the addition in the CMD of the 1103 stars measured in the external field,
to allow for a more direct comparison with the empirical diagram.
} 
\label{simvi}
\end{figure}

Fig. \ref{simbv} compares three representative cases of synthetic CMDs (panels
b, c and d), with the empirical one (panel a) in the $V, B-V$ plane. To better 
appreciate the differences between the various theoretical predictions, in
panels b, c and d we plot only the 837 synthetic stars attributed to Be\,17.
Fig. \ref{simvi} shows the corresponding $V, V-I$ plane, with  the
superposition of the 1103 external field stars to the 837 synthetic ones.
Hence, all the CMDs of Fig. \ref{simvi} contain 1940 objects. 
The observed $V-I$ cluster sequence looks slightly wider than the
synthetic one, perhaps due to the character of the field contamination, but we 
think this does not significantly affect the age determination.
It is also immediately apparent that our synthetic MSs never reach as deep as
the empirical one. This is due to the fact that none of the adopted sets of
stellar evolution models includes stars less massive than 0.6 \msun, while our
photometry clearly detects significantly smaller stars.

The binary sequence is evident in the synthetic CMDs of Fig. \ref{simbv} and it
is interesting to notice that it perfectly overlaps with existing sequences of
the observational diagram, one on the right of the MS and one above the MSTO.
These two observational sequences are not statistically significant by
themselves, due to the field contamination, but the perfect match with the
predicted binary sequence is unlikely to be just a fortuitous coincidence.

With all the adopted sets of stellar evolution models, the age that allows us
to better reproduce all the observed properties of Be\,17's CMD is between 8.5
Gyr (with the BBC models, panel b in Figs \ref{simbv} and {\ref{simvi}) and 9
Gyr (with the FST models, panel d in Figs \ref{simbv} and {\ref{simvi}).
Namely, with this age we obtain the right luminosity and colours of both the
MSTO and the clump, the right shape of the MS, the subgiant branch (SGB) and
the RGB, the right number of stars predicted on the clump and on the RGB, and
the right star counts at the various MS levels. Younger ages imply fainter
clump, worse MSTO morphology and, often, more extended SGB, and are therefore
easy to reject. On the other hand, it is admittedly difficult to completely
rule out older ages, up to 11-12 Gyr. As shown in panel c of Figs \ref{simbv}
and \ref{simvi} (BBC models with Z=0.008, age = 12 Gyr, \ebv = 0.56 and \mmm =
12.1), the clump does become brighter and bluer, the star distribution on the
RGB does differ somehow from the observed one, the SGB luminosity does increase
too much with decreasing temperature, but none of these defects can really be
considered bad enough to reject the case, once both the theoretical and the
observational uncertainties are taken into account. However, since all the
adopted sets of stellar evolution models agree in favouring an age of $9.0 \pm
0.5$ Gyr, we consider this value as the most likely age of Be\,17.

The various sets of stellar models converge to values in strikingly good
agreement also in the predictions for the reddening and the distance modulus.
We obtain \ebv = 0.62, \mmm = 12.2 for the BBC models with Z=0.008 and age =
8.5 Gyr; \ebv = 0.60, \mmm = 12.2 for the FRA models with Z=0.01 and age = 9.0
Gyr; \ebv = 0.60, \mmm = 12.2 for the FST models with Z=0.01 and age = 9.0 Gyr.
For sake of completeness (and because it was suggested by K94) we have also
tested the possibility of some differential reddening affecting this rather
contaminated region. From the comparison of synthetic CMDs computed with
varying amounts of differential reddening, we suggest that, if present, the
absorption variations should be  rather small, with $\Delta$\ebv $\leq\pm$0.03.

The LFs of the three synthetic CMDs shown in Fig. \ref{simvi} are plotted in
Fig. \ref{simlf} (lines) and compared to that of the CMD of Fig. \ref{simvi}a.
The latter is significantly affected by the 1103 external stars, thus making the
comparison of little help in discriminating between different cases. We simply
find that all these models are consistent with the data, except for the lack of
very low mass stars which lead us to under predict the faint end of the LF.

\begin{figure}
\centering
\includegraphics[bb=110 200 450 400,clip, scale=0.7]{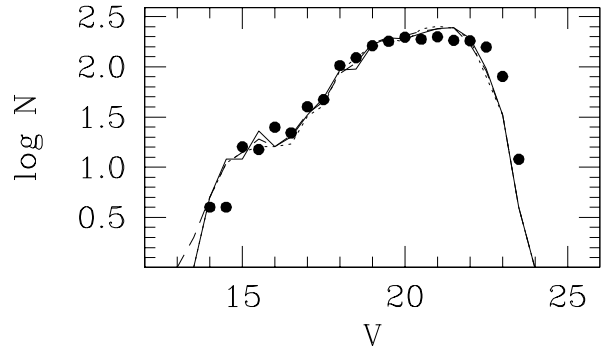} 
\caption{Luminosity functions of the 1940 stars in the cluster region as
measured from the data (dots) and from the synthetic CMDs of Fig. \ref{simvi}.
Error bars on the empirical values are within the size of the symbols.
The solid line corresponds to the objects in panel b, the dotted line to those
in panel c and the dashed line to those in panel d: they are barely
distinguishable from each other.} 
\label{simlf}
\end{figure}

\section{Summary and discussion}

\subsection{Comparison with previous results}

As already mentioned in the Introduction, Be~17 has received a lot of attention 
in the last decade.

The first to present good CMDs for this cluster was K94; he observed the
central part of Be~17 with the 2.1 m telescope and a larger area ($\sim 23$
arcmin$^2$) with the 0.9 m telescope, both at Kitt Peak. He deduced a cluster
diameter of about 13 arcmin and mentioned the possibility of differential
reddening. By comparison of the CMD of the central part to NGC~6791 and
assuming that Be~17 is metal-poorer, he found an age equal or slightly larger
than that of NGC~6791, an apparent distance modulus $(m-M)_V > 15.0$, and a
reddening $E(V-I) > 0.7$ (or $E(B-V) > 0.56 $). He also concluded that the
field stars appearing redder and brighter than the cluster MS belong to the
Perseus spiral arm.  The results for age and reddening are in very good
agreement with  our findings. We find instead a much shorter distance, but
cannot readily explain such a difference.

The same year \cite{jp94}, in their study of the properties of the old OCs,
using photometry presented by \cite{pjm94} and their
calibration of the MAI, derived the very large age of 12 Gyr for Be~17. The MAI
had not been intended to measure absolute ages, and \cite{jp94} stress that it
is to be used only to measure relative ages. In their calibration of the
$\delta V$ versus age (the latter taken from internally precise, but
inhomogeneous sources) they also considered several GCs, with ages taken from
\cite{chaboyer92}. Among them the youngest GCs have ages of about 10 Gyr. The
MAI was  scaled so that an age of 15 Gyr was obtained for the oldest among the
globulars: an age currently considered older than the Universe !

Another kind of analysis was done by P97, who concentrated on the single
cluster, instead of considering the bulk properties of the open cluster sample.
He obtained deep and precise $B, V, I$ photometry of Be~17, using the 2.1 m 
Kitt Peak telescope on a photometric night. P97 noted, as we already said in
Sect. 2.1, that his and K94 photometries do not perfectly agree, but he thought
that this did not compromise his major conclusions. He used a control field to
statistically subtract the field star contamination and employed the cleaned
CMDs for his analysis. First he determined the cluster age by means of the
$\delta V$ and the $\delta 1$ (this is a difference in colour, that can be
calibrated to derive a second estimate of $\delta V$): he found $\delta V = 2.7$
(instead of 2.8 as in \citealt{pjm94})
and from the MAI calibration obtained an age of 10.9 Gyr. 
Later he derived the cluster age from isochrone fitting. Using the \cite{van85}
isochrones (that do not reach the RC) he derived a best fit age of 12 Gyr with
Z=0.01, while for the \cite{bertelli94} isochrones he cited a range of 10 to 12
Gyr, with a preference for Z=0.02. On the whole, he assumed
an age of 12$^{+1}_{-2}$ Gyr and a metallicity between solar and half of solar.
The reddening and distance he derived are perfectly consistent with ours:
$0.52 \le E(B-V) \le 0.68$ and $(m-M)_0 = 12.15 \pm 0.10$.
With these values, and adding the largish diameter and mass and the rather high
radial velocity, P97 seems to think that Be~17 may be a transitional cluster,
in between open and globular ones; his main conclusion is that this cluster
poses important constraints on the age of the disc and of the halo too, which
is thought to have formed before.

Why did he obtain such a large age using the \cite{bertelli94} isochrones,
coming from the same set of models from which we obtain instead a best fit age
of 8.5 Gyr and Z=0.008 ? With his data we would have made a choice different
from his. In fact, judging from his figures 7 and 8, the luminosity of the RC
is actually better fit by the 10 Gyr isochrone. Furthermore, we would have
given lower weight to the \cite{van85} isochrones, since they miss the very
important clue of fitting both the MSTO and the RC luminosities and colours.

A confirmation of our choice comes from the work by \cite{carraro98a}:
they re-analysed the data by P97 (and K94), using the new \cite{girardi00}
isochrones (in preparation, at the time) and found that the possible
combinations of parameters are:
$E(B-V) = 0.55-0.67$, $(m-M)_0 = 12.3 \pm 0.07$, and age$=9 \pm 1$ Gyr for
Z$=0.007 - 0.013$, with the metallicity not affecting the age. They too
differ from P97's choice of the best fit from the \cite{bertelli94} isochrone,  
and place lower weight to the \cite{van85} ones.

Another check of the cluster parameters come from the work by \cite{carraro99}
based on $J, K$ IR data. They obtained an independent estimate of the cluster
metallicity from the RGB slope, finding [Fe/H]$\simeq-0.35$, in very good 
agreement with the spectroscopic estimates. Using this value and the
\cite{girardi00} isochrones they find a best fit age of 9 Gyr; translating from 
the IR quantities, they derive $E(B-V) = 0.58$ and $(m-M)_0 = 12.00 \pm 0.1$.

\cite{salaris04} derived a calibration of the $\delta V$ that takes
into account also the metallicity. They determined distances to 10 clusters by
means of main sequence fitting, and the corresponding ages using their own
evolutionary models. These clusters
were used to derive the calibration from which,
using literature values for $\delta V$  and metallicity,  an age of 10 Gyr is
obtained for Be~17. They also confirm a gap of 2.0$\pm1.5$ Gyr  between the
formation of the thin disc and of the halo, while the oldest OCs (Be~17 and
NGC~6791) have the same age of the oldest thin disc stars.

Finally, \cite{kc06} assumed the metallicity from \cite{friel02} and obtained,
from isochrone fit without RC, an age of 10 $\pm$ 1 Gyr, $(m-M)_V$=14.1-14.2
and a \ebv=0.56-0.61. They do mention that assuming a higher metallicity (as
indicated by \citealt{friel05}) the age would decrease, thus presumably
becoming closer to our value.

Since we have not analysed yet the spectra we obtained for three
RC stars, no real comparison between our (approximate) photometric metallicity
based only on the best fit isochrones and the spectroscopic values obtained by
\cite{friel02} and \cite{friel05} is attempted. We only note that there
is a general agreement that the cluster metallicity is (slightly) sub solar.

\subsection{Discussion}

Our analysis confirms that Be~17 is a very old disc object but that its age
is far from that of globulars.
There is only one known exception: 
\cite{carraro05} has presented evidence that GCs 
may be much younger than usually taken for granted. Whiting~1, with
$b = -60.6^o$, low metallicity (Z = 0.001), and
heliocentric distance of 45 kpc hardly qualifies as a disc open cluster, yet
it has an age of about 5 Gyr. Its nature has to be further investigated;
either the separation between Galactic globular and open clusters
is fuzzier than we think, or Whiting~1 is a "freak", maybe connected
to an accretion event.

Considering the age distribution of the bulk of well behaved globular clusters,
recent studies of a very ample, well defined, homogeneously analysed sample of
GCs (\citealt{rosen99,deangeli05}) have derived accurate relative
ages for 55 GCs, finding that most of them are coeval and old. There are
some GCs younger by about 1 Gyr, but the really younger ones (by about 3 Gyr)
seem all connected to streams, i.e. not truly genuine halo or disc clusters.
The absolute age is more delicate to determine; for instance,
\cite{gratton03}, after deriving a very accurate distance using the main 
sequence fitting method for three  GCs also present in that sample, found that
two of them, NGC~6752 and NGC~6397, are coeval while the metal-richer 47~Tuc is
younger by about 2.5 Gyr. This, using models by \cite{scl97}, 
and taking into account differences in microscopic diffusion and helium content
between those models and later results, means absolute ages of about 13 and
less than 11 Gyr, respectively, with an error bar of about 1 Gyr.

Our result places Be~17 quite safely away from that lower limit; however,  to
obtain an absolute age we have to rely on theoretical models,
and ages of GCs and OCs obtained by different authors with different techniques
and models may not be immediately comparable. To get a definitive answer,
stricter homogeneity is required, but the evidences are still largely in favour
of the thin disc being younger than the halo.

Finally, we note that a  hiatus between the formation of the halo and thick
disc  and that of the presently observed thin disc has also been found using
chemical signatures. For instance,  \cite{gratton00} analysed disc and halo
stars in the solar neighborhood and noticed that the run of [Fe/O] versus [O/H]
supports the hypothesis of a period with no, or very low, star formation,
confirming previous results by \cite{f98}. In a [Fe/O] versus [O/H] plot (where
O is the clock), Fe -
primarily produced by type Ia Supernovae - after remaining constant with
increasing O, suddenly increases at constant O. Since O is
instead produced by short-lived, massive stars, this is interpreted as
absence of such stars, i.e., of star formation, for a period of at least 1
Gyr.

\subsection{Summary}

We have analysed $B,V,I$ photometric data of Be~17, and determined its
fundamental parameters by means of comparison of the observed CMDs to
synthetic ones generated using different sets of stellar evolutionary models.

\begin{itemize}

\item
We find an age of 8.5-9 Gyr, a distance \mmm = 12.2\ (consistently for all
evolutionary tracks used; an error of $\pm$ 0.1 mag can be assumed), a reddening
value \ebv = 0.60-0.62, and approximate metallicity Z=0.008 or 0.01.
The three adopted sets of models agree very well on these values.

\item
We cannot completely rule out a larger age ($\lsim 12$ Gyr) but the
corresponding synthetic CMDs produce consistently worse fits to the observed
ones. 

\item
Differential reddening is estimated to be at most at the 5 per cent level.

\item
We find that a binary fraction of 30 per cent well reproduces the MS: these
binary systems populate a secondary brighter and redder sequence that is
possible to detect also in the observed CMDs along the single-stars MS and
brighter than the MSTO.

\item
The $\delta V$ measured on our data would imply a very old age ($\gsim 10$ Gyr),
but a re calibration of this useful relative age indicator, based on
homogeneously determined ages, is necessary.

\item
Comparison of our findings with literature values indicates only two real
discrepancies: with K94 for distance and with P97 for age; possible causes for
the latter have been discussed.

\item
The comparison field shows a structure probably associated to the young
population of
the Perseus arm; there is a second component, compatible with having
intercepted the orbit of the disrupting CMa galaxy.

\item
Be~17 has a sub solar metallicity, a quite large reddening, and a very old age.
With our preferred value of 8.5 -- 9 Gyr Be~17 is definitely younger than the
bulk of GCs, those that are thought to have formed in the very first phases of
Galaxy formation, either by a monolithic collapse or early fragment accretion.

\end{itemize}

Further work on Be~17 is foreseen, both to derive its elemental abundances on a
scale homogeneous with the other OCs in our sample (e.g., \citealt{carretta05}
and references therein) and to define membership from radial velocities for
about 100 stars in crucial evolutionary phases. The latter information may be
used to confirm or refine the determination of cluster properties done in
the present paper.

\bigskip\noindent
ACKNOWLEDGEMENTS
\par\noindent
We gratefully acknowledge the use of software written by P. Montegriffo,
and of the BDA, operated for many years by J.C. Mermilliod and now
by Dr. E. Paunzen. We thank L. Di
Fabrizio for help with the photometric calibration, and M. Bellazzini
for discussions on CMa. The bulk of the synthetic
CMD code was originally provided by Laura Greggio. We thank the anonymous 
referee for the useful suggestions.
This project has received partial financial support from the Italian MIUR
under PRIN 2003029437.
AB thanks the Fundaci\'on Galileo Galilei for funding a visit to La Palma
during which the analysis of the cluster was completed.

\end{document}